\documentclass{cernrep}
\usepackage{texnames}
\usepackage[T1]{fontenc}
\begin{document}

\title{Testing for a Signal}
\author{Wolfgang A. Rolke and Angel M. L\'{o}pez}
\institute{University of Puerto Rico - Mayaguez}
\maketitle

\begin{abstract}
We describe a statistical hypothesis test for the presence of a signal based
on the likelihood ratio statistic. We derive the test for one case of
interest and also show that for that case the test works very well, even far
out in the tails of the distribution. We also study extensions of the test
to cases where there are multiple channels.
\end{abstract}

\section{Introduction}

In recent years much work has been done on the problem of setting limits in
the presence of nuisance parameters, beginning with the seminal paper by
Feldman and Cousins ~\cite{bib:feldmancousins}. A fairly comprehensive
solution of this problem was given in Rolke, L\'{o}pez and Conrad ~\cite%
{bib:rlc}. In this paper we will study a related problem, namely that of
claiming a new discovery, say of a new particle or decay mode. Statistically
this falls under the heading of hypothesis testing. We will describe a test
derived in a fairly standard way called the likelihood ratio test. The main
contribution of this paper is the study of the performance of this test.
This is essential for two reasons. First, discoveries in high energy physics
require a very small false-positive, that is the probability of falsely
claiming a discovery has to be very small. This probability, in statistics
called the type I error probability $\alpha$, is sometimes required to be as
low as $2.87\cdot10^{-7}$, equivalent to a 5$\sigma$ event. The likelihood
ratio test is an approximate test, and whether the approximation works this
far out in the tails is a question that needs to be investigated. Secondly,
in high energy physics we can often make use of multiple channels, which
means we have problems with as many as 30 parameters, 20 of which are
nuisance parameters. The sizes of the samples needed to insure that the
likelihood ratio test works need to be determined.

\section{Likelihood Ratio Test}

We will consider the following general problem: we have data $\mathbf{X}$
from a distribution with density $f(\mathbf{x};\theta )$ where $\theta $\ is
a vector of parameters with $\theta \in \Theta $ and $\Theta $ is the entire
parameter space. We wish to test the null hypothesis $H_{0}:\theta \in
\Theta _{0}$ (no signal) vs the alternative hypothesis. $H_{a}:\theta \in
\Theta _{0}^{c}$\ (some signal), where $\Theta _{0}$ is some subset of $%
\Theta $. The likelihood function is defined by%
\[
L(\theta |\mathbf{x})=f(\mathbf{x};\theta )
\]%
and the likelihood ratio test statistic is defined by%
\[
\lambda (\mathbf{x})=\frac{\sup_{\Theta _{0}}L(\theta |\mathbf{x})}{%
\sup_{\Theta }L(\theta |\mathbf{x})}
\]%
Intuitively we can understand the statistic in the case of a discrete random
variable. In this case the numerator is the maximum probability of the
observed sample if the maximum is taken over all parameters allowed under
the null hypothesis. In the denominator we take the maximum over all
possible values of the parameter. The ratio of these is small if there are
parameter points in the alternative hypothesis for which the observed sample
is much more likely than for any parameter point in the null hypothesis. In
that case we should reject the null hypothesis. Therefore we
define the likelihood ratio test to be: reject the null hypothesis if $\lambda
(\mathbf{x})\leq c$, for some suitably chosen $c$, which in turn depends on
the type I error probability $\alpha $.

How do we find $c$? For this we will use the following theorem: under some
mild regularity conditions if $\theta \in \Theta _{0}$ then $-2\log \lambda (%
\mathbf{x})$ has a chi-square distribution as the sample size $n\rightarrow
\infty $. The degrees of freedom of the chi-square distribution is the
difference between the number of free parameters specified by $\theta \in
\Theta _{0}$ and the number of free parameters specified by $\theta \in
\Theta $.

A proof of this theorem is given in Stuart, Ord and Arnold ~\cite{bib:stuart}
and a nice discussion with examples can be found in Casella and Berger ~\cite%
{bib:casella}.

\section{A Specific Example: A Counting Experiment with Background and
Efficiency}

We begin with a very common type of situation in high energy physics
experiments. After suitably chosen cuts we find $n$ events in the signal
region, some of which may be signal events. We can model $n$ as a random
variable $N$ with a Poisson distribution with rate $es+b$ where $b$ is the
background rate, $s$ the signal rate and $e$ the efficiency on the signal.
We also have an independent measurement $y$ of the background rate, either
from data sidebands or from Monte Carlo and we can model $y$ as a Poisson
with rate $\tau b$, where $\tau$ is the relative size of the sidebands to
the signal region or the relative size of the Monte Carlo sample to the data
sample, so that $y/\tau $ is the point estimate of the background rate in
the signal region. Finally we have an independent measurement of the
efficiency $z$, usually from Monte Carlo, and we will model $z$ as a
Gaussian with mean $e$ and standard deviation $\sigma _{e}$. So we have the
following probability model:%
\[
N\sim Pois(es+b)\qquad Y\sim Pois(\tau b)\qquad Z\sim N(e,\sigma _{e})
\]%
In this model $s$ is the parameter of interest, $e$ and $b$ are nuisance
parameters and $\tau $ and $\sigma _{e}$ are assumed to be known. Now the
joint density of $N$, $Y$ and $Z$ is given by%
\[
f(n,y,z;e,s,b)=\frac{\left( es+b\right) ^{n}}{n!}e^{-\left( es +b\right) }%
\frac{\left( \tau b\right) ^{y}}{y!}e^{-\tau b}\frac{1}{\sqrt{2\pi \sigma
_{e}^{2}}}e^{-\frac{1}{2}\frac{(z-e)^{2}}{\sigma _{e}^{2}}}
\]%
Finding the denominator of the likelihood ratio test statistic $\lambda $
means finding the maximum likelihood estimators of $e, s ,b$. They are given
by $\widehat{s }=n-y/\tau $, $\widehat{b}=y/\tau $ and $\widehat{e}=z$.

We wish to test $H_{0}: s =0$ vs $H_{a}: s >0$, so under the null hypothesis
we have%
\[
\begin{tabular}{l}
$\log f(n,y,z;0,b,e)=n\log \left( b\right) -\log (n!)-b+$ \\ 
$y\log (\tau b)-\log (y!)-(\tau b)-\frac{1}{2}\log (2\pi \sigma _{e}^{2})-%
\frac{1}{2}\frac{(z-e)^{2}}{\sigma _{e}^{2}}$%
\end{tabular}%
\]%
and we find that this is maximized for $\widetilde{b}=\frac{n+y}{1+\tau }$
and $\widetilde{e}=z$. Now 
\[
\begin{tabular}{l}
${\ \lambda (n,y,z)=}\frac{\sup {\ L(0,b,e|n,y,z)}}{\sup {\ L(s ,b,e|n,y,z)}}%
{\ =}\frac{{\ f(n,y,z|0,}\widetilde{{\ b}}{\ ,}\widetilde{{\ e}}{\ )}}{{\
f(n,y,z|}\widehat{{\ s }}{\ ,}\widehat{{\ b}}{\ ,}\widehat{{\ e}}{\ )}}{\ =} 
$ \\ 
$\frac{\left( \frac{{\ n+y}}{{\ 1+\tau }}\right) {\ /n!}{\ \exp }{\ (-}\frac{%
{\ n+y}}{{\ 1+\tau }}{\ )}\left( {\ \tau }\frac{{\ n+y}}{{\ 1+\tau }}\right)
^{{\ y}}{\ /y!}{\ \exp }{\ (-}{\ \tau }\frac{{\ n+y}}{{\ 1+\tau }}{\ )}\frac{%
1}{\sqrt{{\ 2\pi \sigma }_{{\ e}}^{{\ 2}}}}{\ e}^{{\ -}\frac{{\ 1}}{{\ 2}}%
\frac{({\ z-z})^{{\ 2}}}{{\ \sigma }_{{\ e}}^{{\ 2}}}}}{{\ n}^{{\ n}}{\
/n!\exp (-n)}{\ y}^{{\ y}}{\ /y!\exp (-y)}\frac{1}{\sqrt{{\ 2\pi \sigma }_{{%
\ e}}^{{\ 2}}}}{\ e}^{{\ -}\frac{{\ 1}}{{\ 2}}\frac{({\ z-z})^{{\ 2}}}{{\
\sigma }_{{\ e}}^{{\ 2}}}}}{\ =}$ \\ 
$\frac{\left( \frac{{\ n+y}}{{\ 1+\tau }}\right) ^{{\ n+y}}{\ \tau }^{{\ y}}%
}{{\ n}^{{\ n}}{\ y}^{{\ y}}}$%
\end{tabular}
\]
One special case of this needs to be do be studied separately, namely the
case $y=0$. In this case we can not take the logarithm and the maxima above
have to be found in a different way. It turns out that the MLE's are $%
\widehat{s }=n$, $\widehat{b}=0$ , $\widehat{e}=z$, and under the null
hypothesis we find $\widetilde{b}=\frac{n}{1+\tau }$ and $\widetilde{e}=z$.
With this we find $\lambda (n,0,z)=(1+\tau )^{-n}$.

First we note that the test statistic does no involve $z$, the estimate of
the efficiency. This is actually clear: the efficiency is for the detection
of signal events, but under the null hypothesis there are none. Of course
the efficiency will affect the power curve: if $e$ is small the observed $n$
will be small and it will be much harder to reject the null hypothesis.

Now from the general theory we know that $-2\log \lambda (N,Y,Z)$ has a
chi-square distribution with $1$ degree of freedom because in the general
model there are $3$ free parameters and under the null hypothesis there are $%
2$. So if we denote the test statistic by $L(n,y)$ we get%
\[
\begin{tabular}{l}
$L(n,y)=-2\log \lambda (n,y,z)=$ \\ 
$\left\{ 
\begin{tabular}{ll}
$2\left[ n\log (n)+y\log (y)-(n+y)\log \left( \frac{n+y}{1+\tau }\right)
-y\log (\tau )\right] $ & if $y>0$ \\ 
$2n\log (1+\tau )$ & if $y=0$%
\end{tabular}
\right.$ 
\end{tabular}%
\]
and we have $L(N,Y)\sim \chi _{1}^{2}$, approximately.

Obviously we will only claim a disovery if there is an excess of events in
the signal region, and so the test becomes: reject $H_{0}$ if $n>y/\tau$ and 
$L(n,y)>c$. Now it can be shown that $c=q\chi _{1}^{2}(1-2\alpha )$, the $%
(1-2\alpha)$ quantile of a chi-squared distribution with one degree of
freedom.

The situation described here has previously been studied in Rolke, L\'{o}pez
and Conrad ~\cite{bib:rlc} in the context of setting limits. They proposed a
solution based on the profile likelihood. This solution is closely related
to the test described here. In fact it is the confidence interval one finds
when inverting the test described above.

\section{Multiple Channels}

In high energy physics we can sometimes make use of multiple channels. There
are a number of possible extensions from one channel. We will consider the
following model: there are $k$ channels and we have $N_{i}\sim Pois(e_{i}s_{i}+b_{i})$, $Y_{i}\sim Pois(\tau _{i}b_{i}),$ $i=1,..,k$, all independent. We will again find that the efficiencies do not
affect the type I error probability.
We will discuss two ways to extend the methods above to multiple channels,
both with certain advantages and disadvantages.

\subsection{Method 1: (Full LRT)}

We can calculate the likelihood ratio statistic for the full model. It
turns out that the test statistic $L_{k}$ is given by%
\[
L_{k}(\mathbf{n,y})=\sum_{i=1}^{k}L(n_{i},y_{i})I(n_{i}>y_{i}/\tau _{i})
\]%
where $I$ is the indicator function, that is $I(n>y/\tau)=1$ if $n>y/\tau$,
and $0$ otherwise. In other words the test statistic is simply the sum of 
the test statistics for each channel separately.
The test is then as follows: we reject $H_{0}$ if $L_{k}(\mathbf{n,y})>c$.
It can be shown that the distribution of the test statistic under the null
hypothesis is a linear combination of chi-square distributions. Tables of
critical values as well as a routine for calculating them are available from
the authors.

\subsection{Method 2: (Max LRT)}

Here we will use the following test: reject $H_{0}$ if $M=\max_{i}%
\{L(n_{i},y_{i})I(n_{i}>y_{i}/\tau _{i}\}>c$, that is, we claim a discovery
if there is a significant excess of events in any one channel. For this
method the critical value $c$ is found using Bonferroni's method. We
therefore reject $H_{0}$ if $M>c$, where $c=q\chi _{1}^{2}(1-2(1-\sqrt[k]{%
1-\alpha }))$.

As we shall see soon, which of these two methods performs better depends on
the experiment.

\section{Performance}

How do the above tests perform? In order to be a proper test they first of
all have to achieve the nominal type I error probability $\alpha $. If they
do we can then further study their performance by considering their power
function $\beta (s )$ given by%
\[
\beta (s )=P(\text{reject }H_{0}|\text{ true signal rate is }s )
\]%
Of course we have $\alpha =\beta (0)$. $\beta (s )$ gives us the discovery
potential, that is the probability of correctly claiming a discovery if the
true signal rate is $s >0$.

In simple cases the true type I error probability $\alpha$ and the power $%
\beta (s )$ can be calculated explicitly, in more difficult cases we
generally need to use Monte Carlo. Moreover, if Monte Carlo is used a
technique called importance sampling makes it possible to find the true type
I error probability even out at $5\sigma$.

First we will study the true type I error probability as a function of the
background rate. In figure $1$ we calculate $\alpha $ (expressed in sigma's)
for background rates ranging from $b=5$ to $b=50.$ Here we have used $\tau =1
$ and $\alpha $ corresponding to $3\sigma ,$ $4\sigma $ and $5\sigma $.

\begin{figure}[tbp]
\centering\includegraphics[width=.9\linewidth]{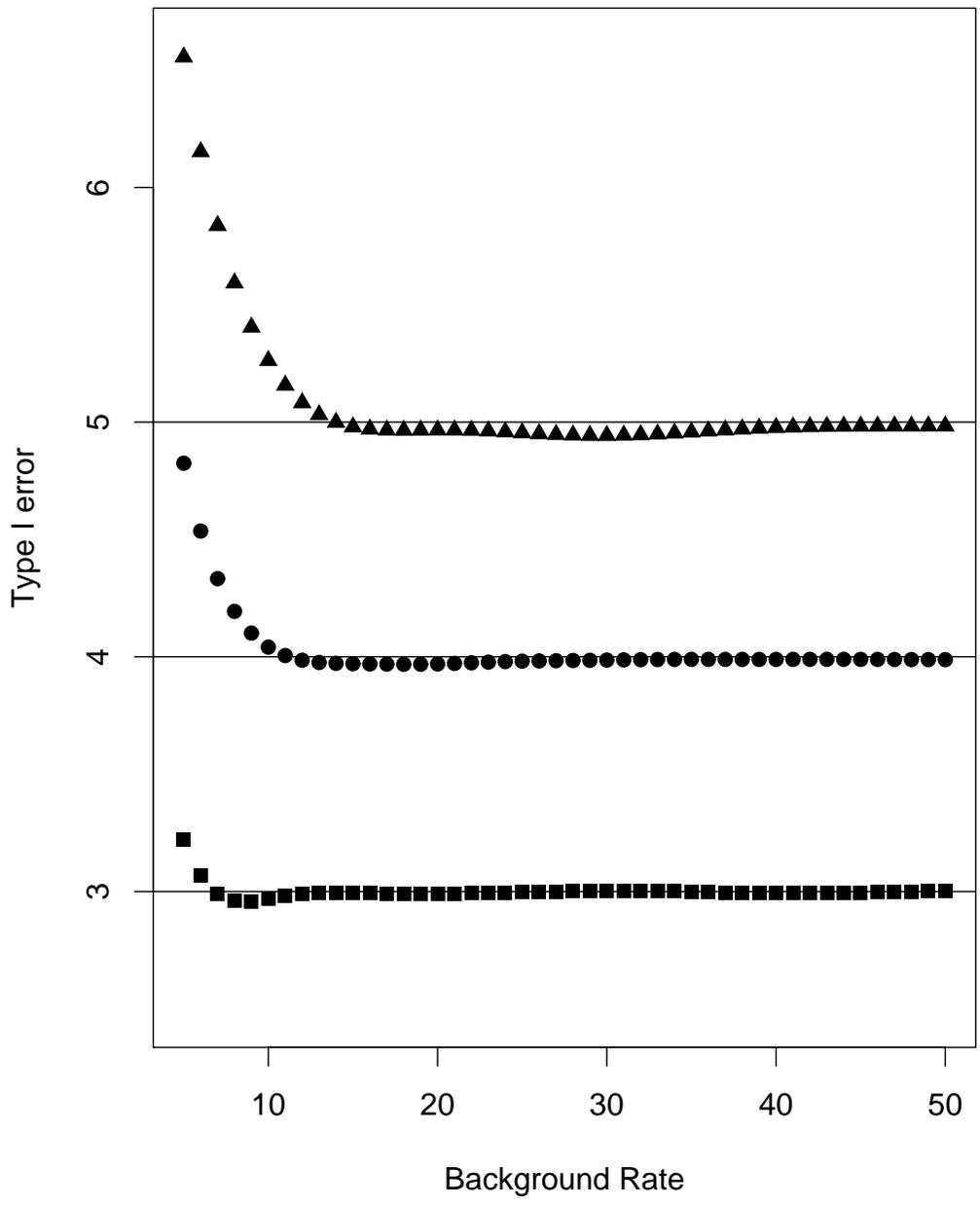}
\caption{Type I error probability $\protect\alpha$ for different values of
the background rate $b$}
\label{fig:rolkelopez_figure1}
\end{figure}

It is clear that even for moderate background rates (say $b>20$) the true
type I error is basically the same as the nominal one. 
For smaller background rates, the method is conservative, that is, the true significance of a 
signal is actually even higher than the one claimed, and it is therefore safe to use 
the method even for small {b}.

In figure $2$ we have the power curves for $b=50$, $\tau =1$, $e=1$, $s $ from $0 $
to $100$ and $\alpha $ corresponding to $3\sigma ,$ $4\sigma $ and $5\sigma $%
. This clearly shows the "penalty" of requiring a discovery threshold of $%
5\sigma $: at that level the true signal rate has to be $83$ for a $90\%$
chance of making a discovery. If $3\sigma $ is used a rate of $52$ is
sufficient, and for $4\sigma $ it is $67$.

\begin{figure}[tbp]
\centering  \includegraphics[width=.9\linewidth]{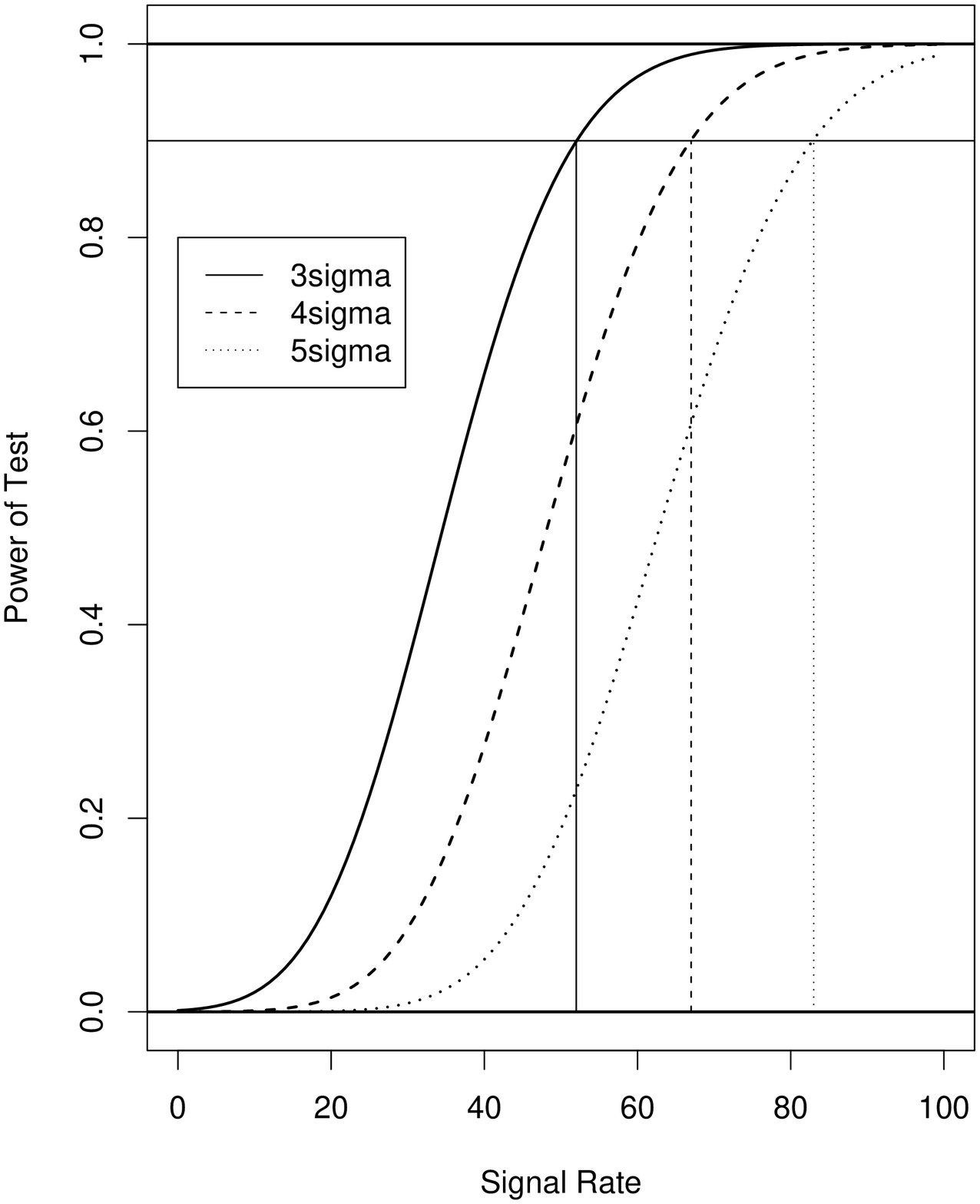}  
\caption{Power of Test for $b=50, \protect\tau=1$}
\label{fig:rolkelopez_figure2}
\end{figure}

Let us now consider the case of multiple channels. In figure $3$ we have the
results of the following simulation: There are $5$ channels, all with the
same background, going from 10 to 100, and the same $\tau=1$. Again we see that 
the test achieves the nominal $\alpha$ even for small background rates.

\begin{figure}[tbp]
\centering  \includegraphics[width=.9\linewidth]{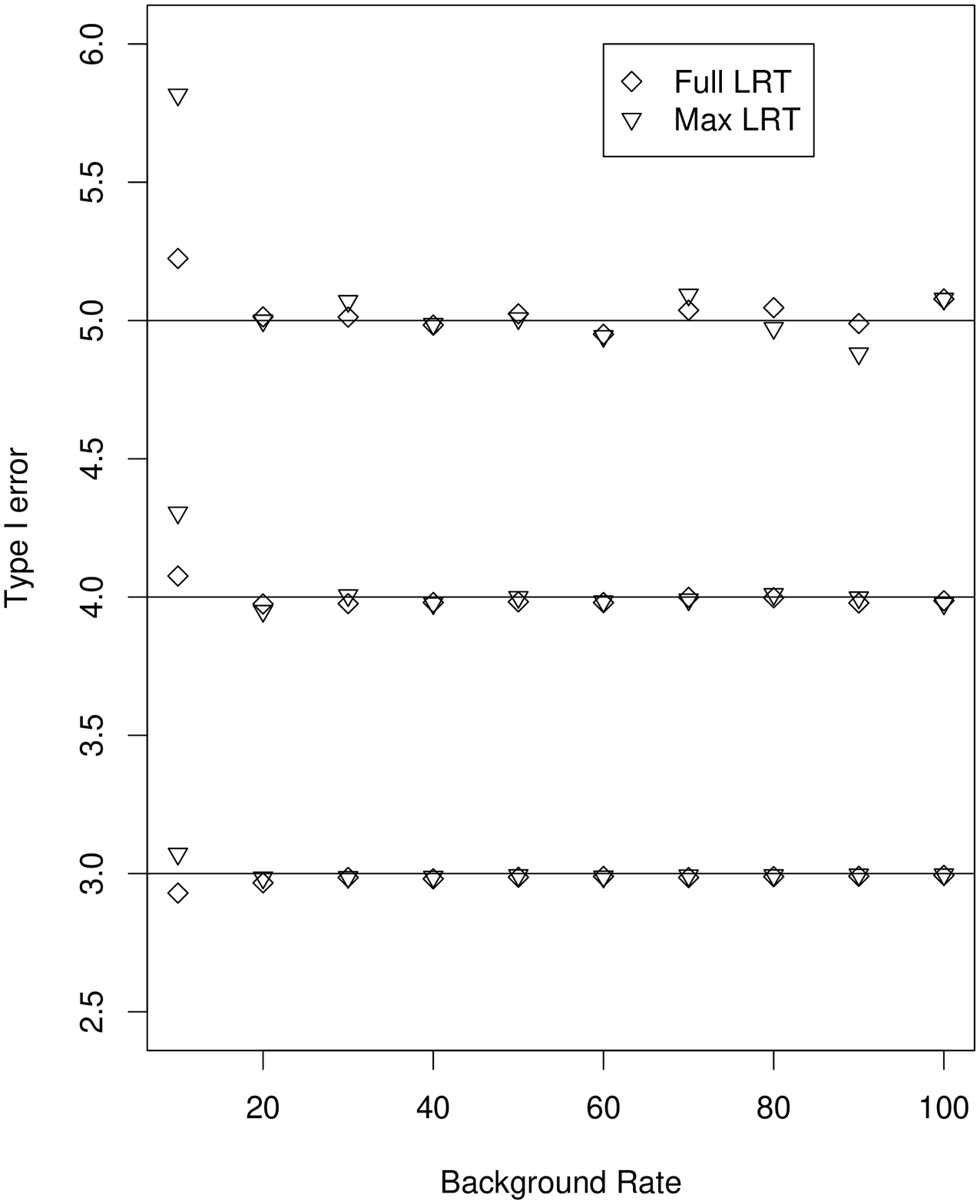}  
\caption{Type I error probability $\protect\alpha$ for different values of
the background rate $b$ for the 5 channel case.}
\label{fig:rolkelopez_figure3}
\end{figure}

For the last study we will compare the two methods for multiple channels. In
figure $4$ we have the power curves for the following situations: we have $5$
channels with $b=50$, $e=1$, and $\tau=1$ for all channels. In case 1 the signal
rate $s$ goes from $0$ to $75$ and is the same in all channels. In case 2 we
have $s_{1}$ going from $0$ to $100$ and $s_{2}=..=s_{5}=0$. All simulations
are done using $\alpha=5\sigma$. Clearly in case 1 Full Lrt does better
whereas in case 2 it is Max Lrt.

\begin{figure}[tbp]
\centering  \includegraphics[width=.9\linewidth]{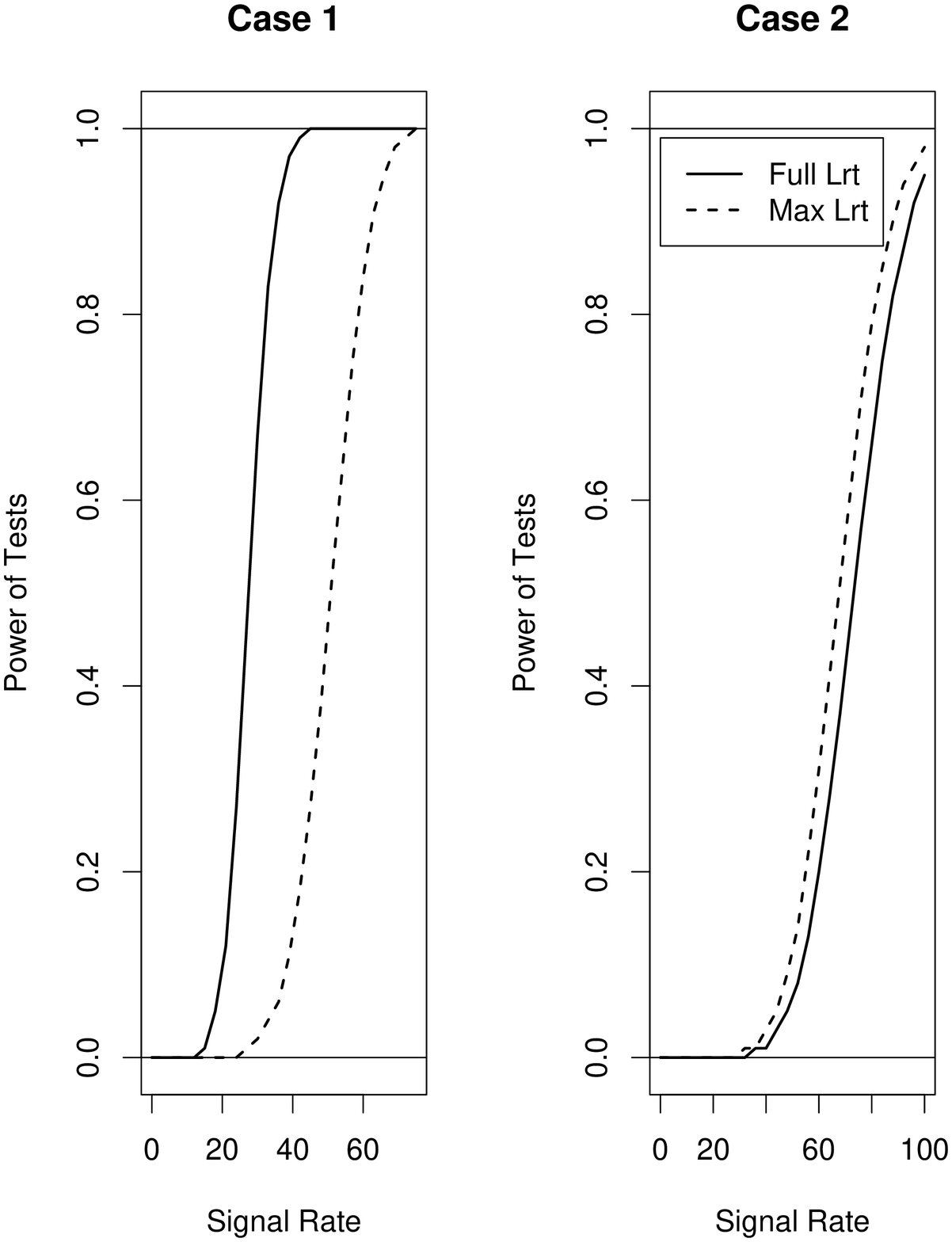}  
\caption{Power of two methods with 5 channels. Case 1 has equal signal in
all channels, case two has signal in channel one and no signals in the
others.}
\label{fig:rolkelopez_figure4}
\end{figure}

This is not surprising because the maximum makes this method more sensitive
to the "strongest" channel whereas the sum makes Full Lrt more sensitive to
a "balance" of the channels. In practice, of course, a decision on which method
to use has to be made before any data is seen. 
A discussion of the optimum strategy for making such a decision is
beyond the scope of this paper.

\section{Further Extensions}

Our extension to multiple channels assumes
possibly different signal rates in each channel.  The most common situation involves different decay channels of a particle whose
existence is being tested.  In that case, the different signal rates
are due to different branching ratios such that $s{i} = r{i}s$ with
a common s.  A detailed discussion of this case along with the inclusion of information on certain variables in each event 
(a technique generally known as marked Poisson) will be found in an
upcoming paper.

\section{Summary}

We have discussed a hypothesis test for the presence of a signal. For the
case of a Poisson distributed signal with a background that has either a
Poisson or a Gaussian distribution we have carried out the calculations and
done an extensive performance study. We have shown that the test achieves
the nominal type I error probability $\alpha $, even at a $5\sigma $ level.
We extended the test to the case of multiple channels with two possible
tests and showed that both achieve the nominal $\alpha $. Either one or the
other has better performance depending on the specific experiment.


\begin{thebibliography}{9}
\bibitem{bib:feldmancousins} R.D. Cousins, G.J. Feldman, \textquotedblleft A
Unified Approach to the Classical Statistical Analysis of Small
Signals\textquotedblright , \textit{Phys. Rev}, \textbf{D57}, (1998) 3873.

\bibitem{bib:rlc} W.A. Rolke, A. L\'{o}pez and J. Conrad, \textquotedblleft
Limits and Confidence Intervals in the Presence of Nuisance
Parameters\textquotedblright , Nuclear Instruments and Methods A, 551/2-3,
2005, pp. 493-503, physics/0403059

\bibitem{bib:stuart} A. Stuart, J.K.Ord and S. Arnold, \textquotedblleft 
\textit{Advanced Theory of Statistics, Volume 2A: Classical Inference and
the Linear Model}\textquotedblright , 6$^{th}$ Ed., London Oxford University
Press (1999)

\bibitem{bib:casella} G. Casella and R.L. Berger, \textquotedblleft \textit{%
Statistical Inference}\textquotedblright , 2$^{nd}$ Ed., Duxbery Press,
(2002)
\end{thebibliography}
\end{document}